\documentclass{PoS}

\graphicspath{{./fig/}}

\title{Seven years of Tunka-Rex operation}

\ShortTitle{Seven years of Tunka-Rex operation}

\author{
\speaker{D.~Kostunin}$^{1}$,
P.~Bezyazeekov$^{2}$,
N.~Budnev$^{2}$,
O.~Fedorov$^{2}$,
O.~Gress$^{2}$,
O.~Grishin$^{2}$,
A.~Haungs$^{3}$,
T.~Huege$^{3,4}$,
Y.~Kazarina$^{2}$,
M.~Kleifges$^{5}$,
E.~Korosteleva$^{6}$,
L.~Kuzmichev$^{6}$,
V.~Lenok$^{3}$,
N.~Lubsandorzhiev$^{6}$,
S.~Malakhov$^{2}$,
T.~Marshalkina$^{2}$,
R.~Monkhoev$^{2}$,
E.~Osipova$^{6}$,
A.~Pakhorukov$^{2}$,
L.~Pankov$^{2}$,
V.~Prosin$^{6}$,
F.~G.~Schr\"oder$^{3,7}$,
D.~Shipilov$^{2}$,
A.~Zagorodnikov$^{2}$
-- Tunka-Rex Collaboration
~\\
$^{1}$DESY, Zeuthen, 15738 Germany\\
$^{2}$Applied Physics Institute ISU, Irkutsk, 664020 Russia\\
$^{3}$Institut f\"ur Kernphysik, Karlsruhe Institute of Technology (KIT), Karlsruhe, 76021 Germany\\
$^{4}$Astrophysical Institute, Vrije Universiteit Brussel, Pleinlaan 2, 1050 Brussels, Belgium\\
$^{5}$Institut f\"ur Prozessdatenverarbeitung und Elektronik, Karlsruhe Institute of Technology (KIT), Karlsruhe, 76021 Germany\\
$^{6}$Skobeltsyn Institute of Nuclear Physics MSU, Moscow, 119991 Russia\\
$^{7}$Bartol Research Institute, Department of Physics and Astronomy, University of Delaware, Newark, DE, 19716, USA
~\\
E-mail: \email{kostunin@tunkarex.info}
}

\abstract{
The Tunka Radio Extension (Tunka-Rex) is a digital antenna array located in the Tunka Valley in Siberia, 
which measures the radio emission of cosmic-ray air-showers with energies up to EeV. 
Tunka-Rex is externally triggered by the Tunka-133 air-Cherenkov timing array (during nights) and by the Tunka-Grande array of particle detectors (remaining time). 
These three arrays comprise the cosmic-ray extension of the Tunka Advanced Instrument for cosmic rays and Gamma Astronomy (TAIGA).
The configuration and analysis pipeline of Tunka-Rex have significantly changed over its runtime.
Density of the antennas was tripled and the pipeline has become more developed forming now sophisticated piece of reconstruction software.
During its lifecycle Tunka-Rex has demonstrated that a cost-effective and full duty-cycle radio detector
can reconstruct the energy and shower maximum with a precision comparable to optical detectors. 
Moreover, it was shown that cosmic-ray instruments, that use different detection techniques and are placed in different locations, can be cross-calibrated via their radio extensions.
These results show the prospects of application of the radio technique for future large-scale experiments for cosmic-ray and neutrino detection. 
For the time being Tunka-Rex has ceased active measurements and focuses on the data analysis and publication of corresponding software and data in an open-access data center with online analysis features. 
In this report we present the current status of the array and give an overview of the results achieved during these years as well as discuss upcoming improvements in instrumentation and data analysis, which can be applied for the future radio arrays.
}

\FullConference{36th International Cosmic Ray Conference -ICRC2019-\\
		July 24th - August 1st, 2019\\
		Madison, WI, U.S.A.}

\begin{document}

\section{Introduction}

The era of multi-messenger astronomy~\cite{Kowalski_Bartos_mm} has began recently and the 
digital radio arrays are a very feasible instrument for measuring ultra-high energy astrophysical messengers in PeV-EeV range~\cite{Kostunin:2019ahp}.
While the first and second generation instruments were focused on technological developments and cosmic-ray measurements, the next (third) generation radio arrays will aim on PeV-gamma~\cite{Haungs:2019ylq} and EeV-neutrino~\cite{Alvarez-Muniz:2018bhp} detection.
In this work we overview the seven years of operation, highlight the main results and discuss the future of the Tunka Radio Extension --- an instrument of the second generation, which has proven the feasibility of ultra-high-energy cosmic-ray detection with sparse radio arrays and has obtained several important results boosted the development of this technique.

\begin{figure}[b!]
\centering
\includegraphics[width=1.0\textwidth]{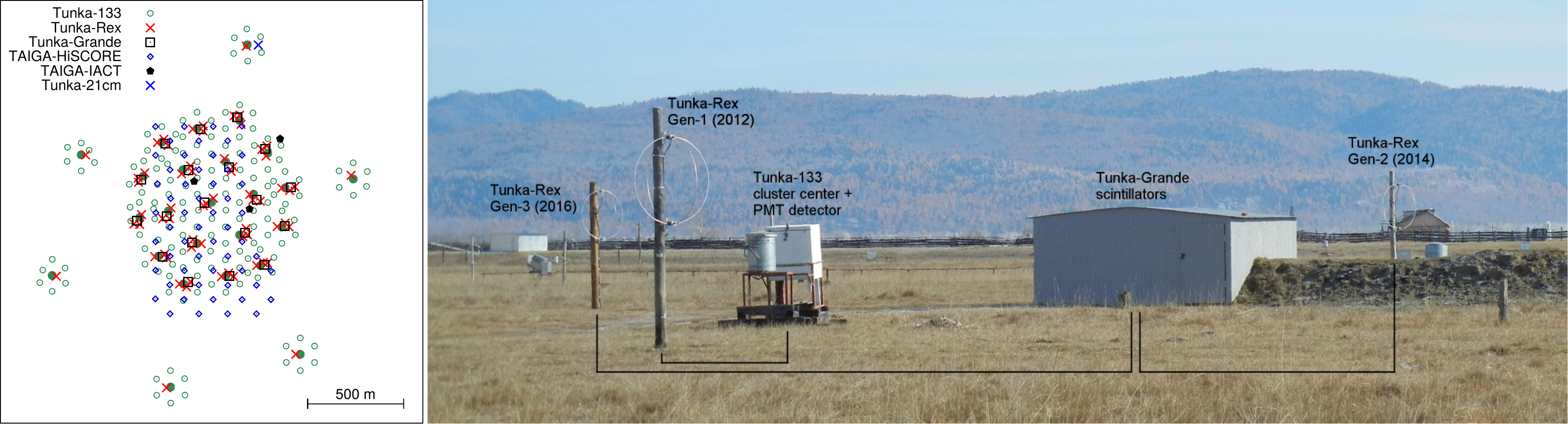}
\caption{\textit{Left:} layout of the TAIGA observatory. Cosmic-ray instruments (Tunka-133, Tunka-Rex and Tunka-Grande) are grouped in clusters.
\textit{Right:} photo of single cosmic-ray cluster of TAIGA observatory. Lines depict the cable connections between Tunka-Rex antennas and DAQ.}
\label{fig:cluster}
\end{figure}

\begin{figure}[b!]
\centering
\includegraphics[width=1.0\textwidth]{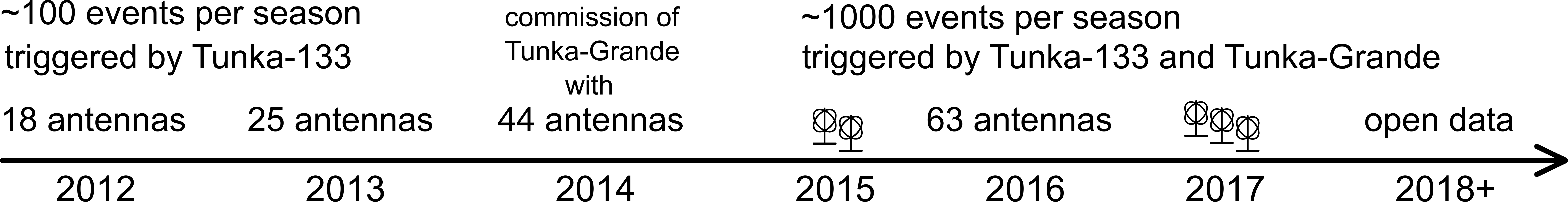}
\caption{Timeline of Tunka-Rex development.
The antenna array has been commissioned in 2012 with 18 antenna stations triggered by the Tunka 133 modules.
Since commission of Tunka-Grande in 2014-2015 Tunka-Rex receives trigger from Tunka-Grande as well (during daytime measurements).
Starting from 2018, we are working on the public access of the Tunka-Rex software and data.}
\label{fig:timeline}
\end{figure}

\section{Detector}
Tunka Radio Extension (Tunka-Rex) is a digital antenna array located at the Tunka Advanced Instrument for cosmic rays and Gamma Astronomy (TAIGA) observatory~\cite{Budnev:2016btu, Kostunin:2019nzy}.
TAIGA instruments can be divided in two main groups: the cosmic-ray instrumentation (Tunka-133~\cite{Prosin:2015voa}, Tunka-Rex~\cite{Bezyazeekov:2015rpa} and Tunka-Grande~\cite{Budnev:2015cha}) and gamma-ray instrumentation (Tunka-HiSCORE~\cite{Tluczykont:2017pin} and TAIGA-IACT~\cite{Yashin:2015lzw}).
In the left side of Fig.~\ref{fig:cluster} one can see the layout of entire observatory and note, that cosmic-ray instruments are grouped in clusters: 19 clusters in a dense core and 6 satellite clusters.
Each core cluster is equipped with 3 Tunka-Rex antenna stations, while satellite clusters contain single antenna stations and no Tunka-Grande scintillators.
In the right side of Fig.~\ref{fig:cluster} one can see a photo of a single cluster with corresponding detectors.

For the time being Tunka-Rex consists of 57 antenna stations located in the dense core of TAIGA (1~km\textsuperscript{2}) and 6 satellite antenna stations expanding the area of array to 3~km\textsuperscript{2}.
Tunka-Rex has been commissioned in 2012 with 18 antenna stations triggered by the air-Cherenkov array Tunka-133.
Each Tunka-Rex antenna station consists of two perpendicular active Short Aperiodic Loaded Loop Antennas (SALLA)~\cite{Abreu:2012pi} with Low Noise Amplifier (LNA).
Signals from the stations are transmitted via 30~m coaxial cables to the analog filter-amplifier, which suppresses the frequencies out of the 30--80~MHz band.
Then the signals are digitized by the local data acquisition system (DAQ) with a 12 bit-sampling at a rate of 200~MHz.
Data are collected in signal traces made of 1024 samples each.
The following years Tunka-Rex has been upgraded several times as well as entire TAIGA observatory, which has been equipped with Tunka-Grande scintillator array providing trigger for Tunka-Rex since 2015.
One can see the timeline of the Tunka-Rex development in Fig.~\ref{fig:timeline}.

\section{Reconstruction of the flux of ultra-high energy cosmic rays}

\begin{figure}[t!]
	\centering
	\includegraphics[height=0.35\linewidth]{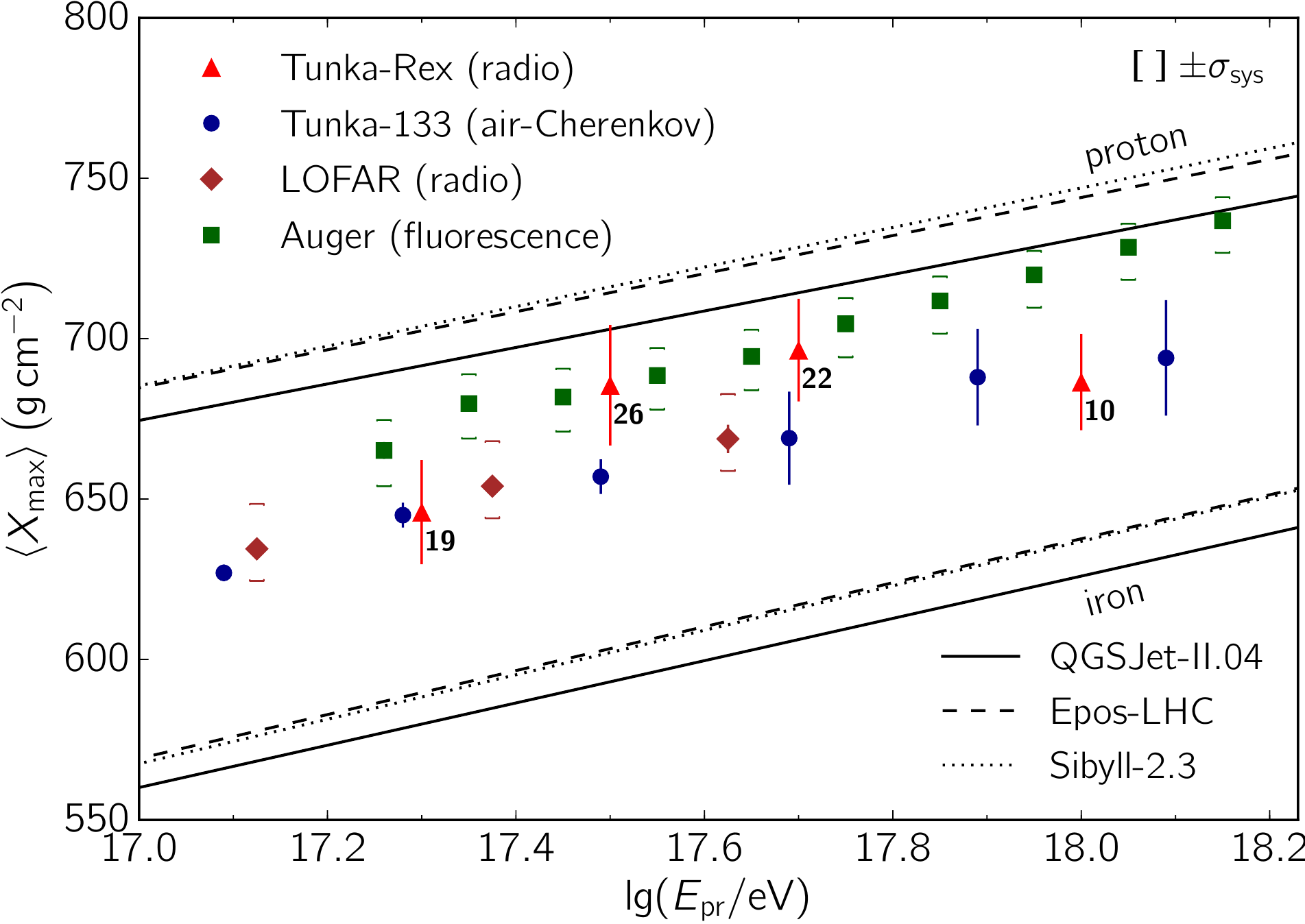}~~~~~
	\includegraphics[height=0.35\linewidth]{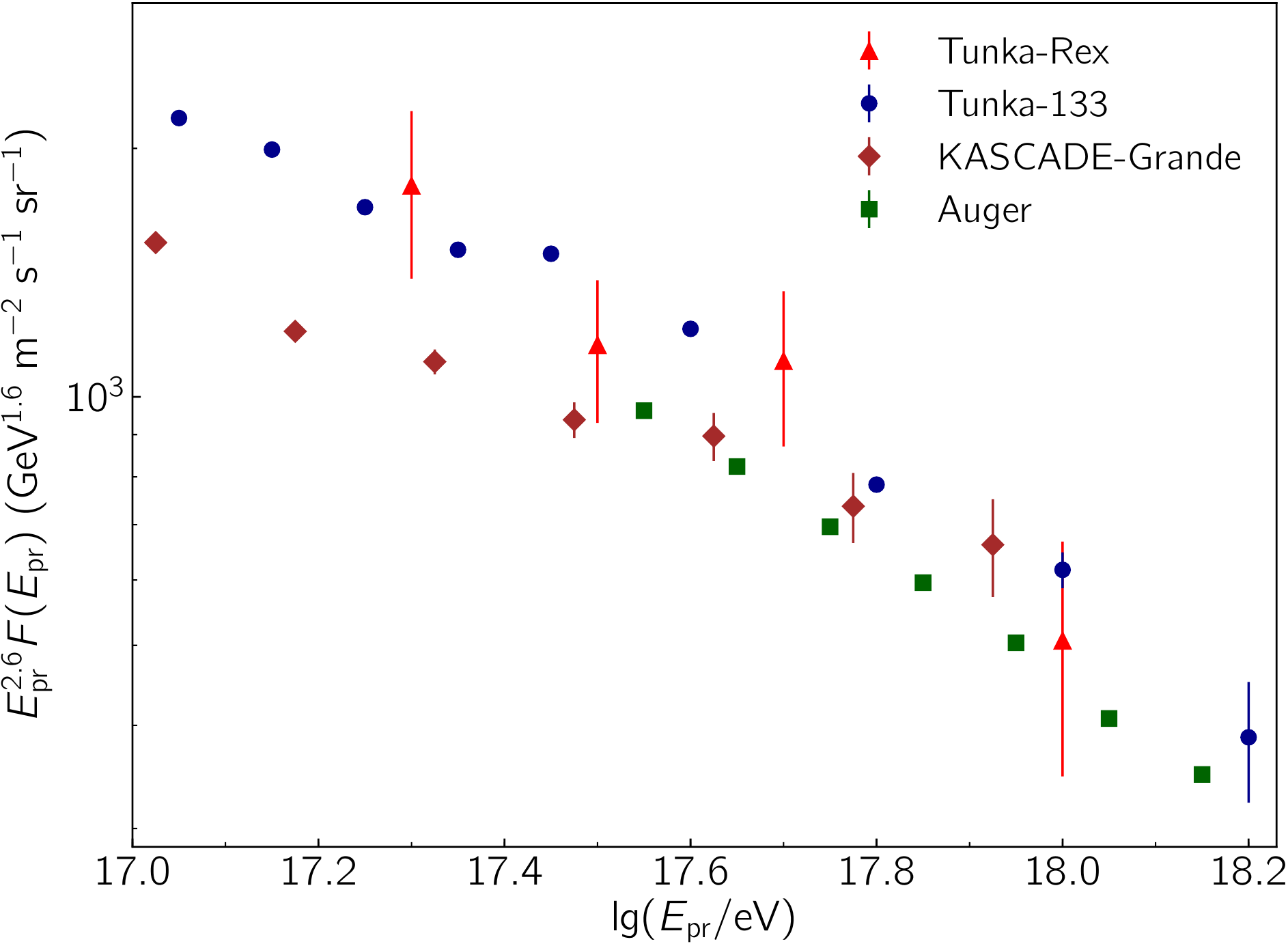}
	\caption{The mean depth of shower maxima \textit{(left)} and the flux of primary cosmic rays \textit{(right)} as a function of primary energy reconstructed from about 10\% of Tunka-Rex data using improved reconstruction methods.
	The values of other experiments are taken from Refs.~\cite{Prosin:2016jev,Buitink:2016nkf,Bellido:2017cgf,Prosin:2016rqu,Bertaina:2015fnz,Fenu:2017hlc}, the model curves are from Refs.~\cite{Ostapchenko:2010vb,Pierog:2006qv,Riehn:2015oba}.
	}	\label{fig:spectra}       
\end{figure}

The original goals of Tunka-Rex were a cross-calibration of the radio and the Cherenkov signal emitted by air showers.
The commission of a radio detector coincided with a release of CoREAS software~\cite{Huege:2013vt}, which is actively used by Tunka-Rex from the very beginning.
Using CoREAS simulations, the methods for the event reconstruction were developed~\cite{Kostunin:2015taa} and tested using semi-blind cross-check with data from Tunka-133~\cite{Bezyazeekov:2015ica}.
A first analysis obtained a reconstruction precision of 15\% for the energy (20\% absolute scale uncertainty) and about 40 g/cm\textsuperscript{2} for the position of the shower maximum. 

After proving the feasibility of radio detection, Tunka-Rex have focused on a precise reconstruction of the primary energy and depth of the shower maximum in a parallel with estimation of the efficiency and exposure of the detector.
Using template fitting based on CoREAS simulations, we have reached the precision for the primary energy of up to 10\% and of up to 25~g/cm\textsuperscript{2} for the depth of the shower maximum~\cite{Bezyazeekov:2018yjw}.
Combining improved reconstruction with the estimation of efficiency and exposure~\cite{Fedorov:2017xih, Lenok:2018das} we have obtained averaged shower maxima and cosmic ray flux as a function of primary energy.
In Fig.~\ref{fig:spectra} one can see the shower maximum and flux spectra produced from about 10\% of the total Tunka-Rex data using these methods.
The progress of the analysis of the rest of the data is given in~\cite{Lenok_ICRC2019}, these proceedings.

Having deeply understood the radio emission phenomena and performance of the detector, it is possible to use the radio instrument for a cross-calibration of the energy scales of independent cosmic-ray detectors located in different environments, equipped with different instruments, etc.
Jointly with LOPES at KASCADE-Grande, Tunka-Rex has successfully proven the agreement between latter and Tunka-133 detectors confirming the consistency of their energy scales on the 10\% level~\cite{Apel:2016gws}.

\section{Tunka-Rex Virtual Observatory}
Following the approach chosen in the German-Russian Astroparticle Data Life Cycle initiative (GRADLCI)~\cite{Bychkov:2018zre} based on the ideas of KDCD~\cite{Haungs:2018xpw}
we are preparing to publish the data of Tunka-Rex under a free data license in the frame of the Tunka-Rex Virtual Observatory (TRVO).
We plan to publish all possible data layers (\texttt{DL}) starting from the lower ones (\texttt{DL0-2}), namely raw radio traces, voltages in channels and reconstructed electrical fields, and finishing with high-level cosmic ray reconstruction and different radio maps (\texttt{DL3+}).
The open data provided by the TRVO can be used by community for the studies of the radio background in the frequency band of 30--80~MHz, searching for radio transients, training of neural networks for RFI tagging, outreach and education, etc.~\cite{Bezyazeekov:2019onw}

We have already deployed several testing databases with subsets of Tunka-Rex events on the servers of the Irkutsk State University and the Karlsruhe Institute of Technology.
The expected number of entries in the database from several data releases is in the order of billions which result in TiB scale of DB.
Currently we are testing the performance of the database and implementing a user interface and basic features\footnote{Already published Tunka-Rex software and datasets can be found on \href{http://soft.tunkarex.info}{http://soft.tunkarex.info}, \href{http://bitbucket.org/tunka/}{http://bitbucket.org/tunka/}}.
Although the data of Tunka-Rex is still under analysis, we will not apply any embargo on it and publish everything as soon as the TRVO framework will be ready for release.

\begin{figure}[t!]
	\centering
	\includegraphics[height=0.35\linewidth]{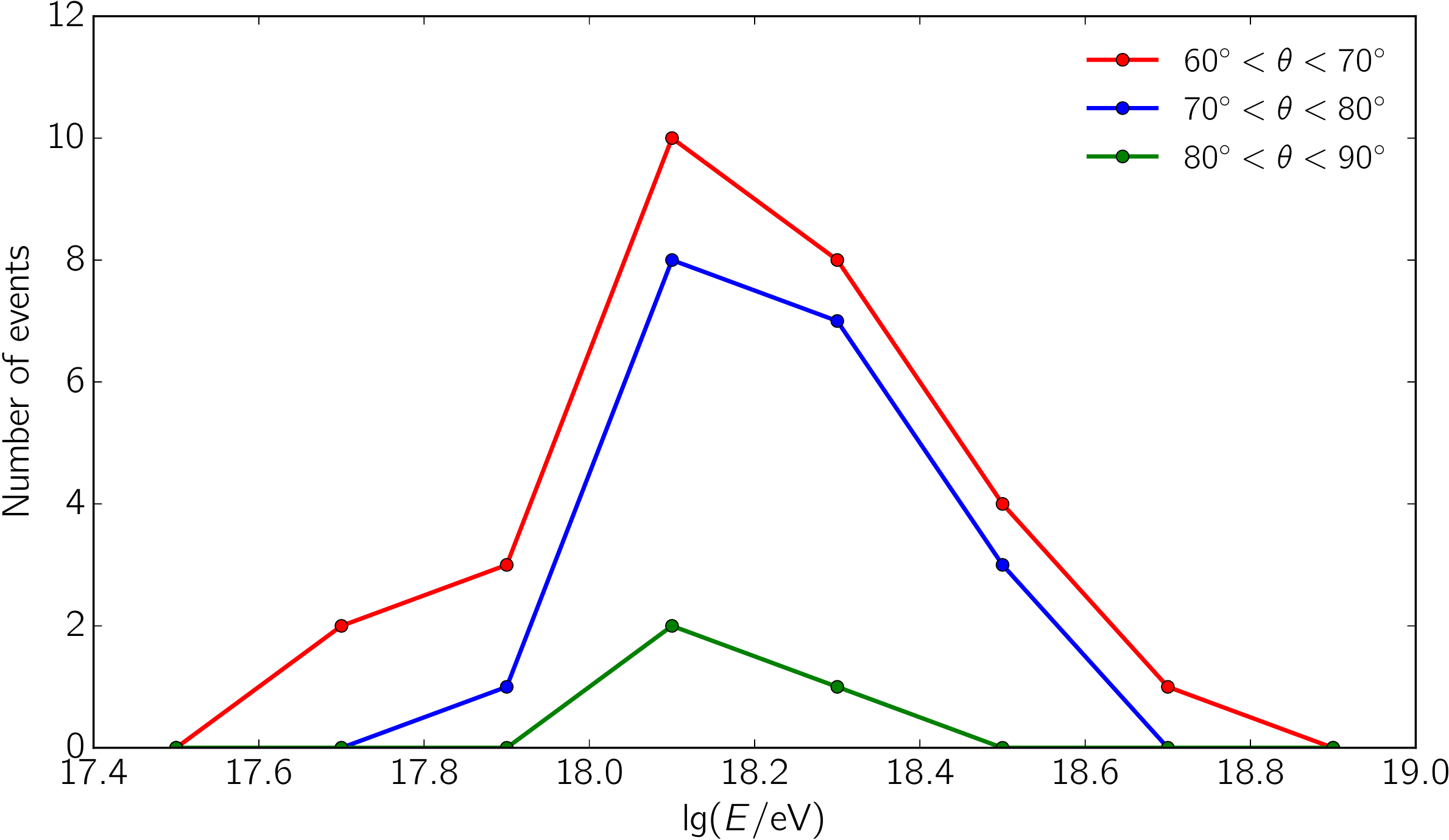}~~~~~
	\includegraphics[height=0.35\linewidth]{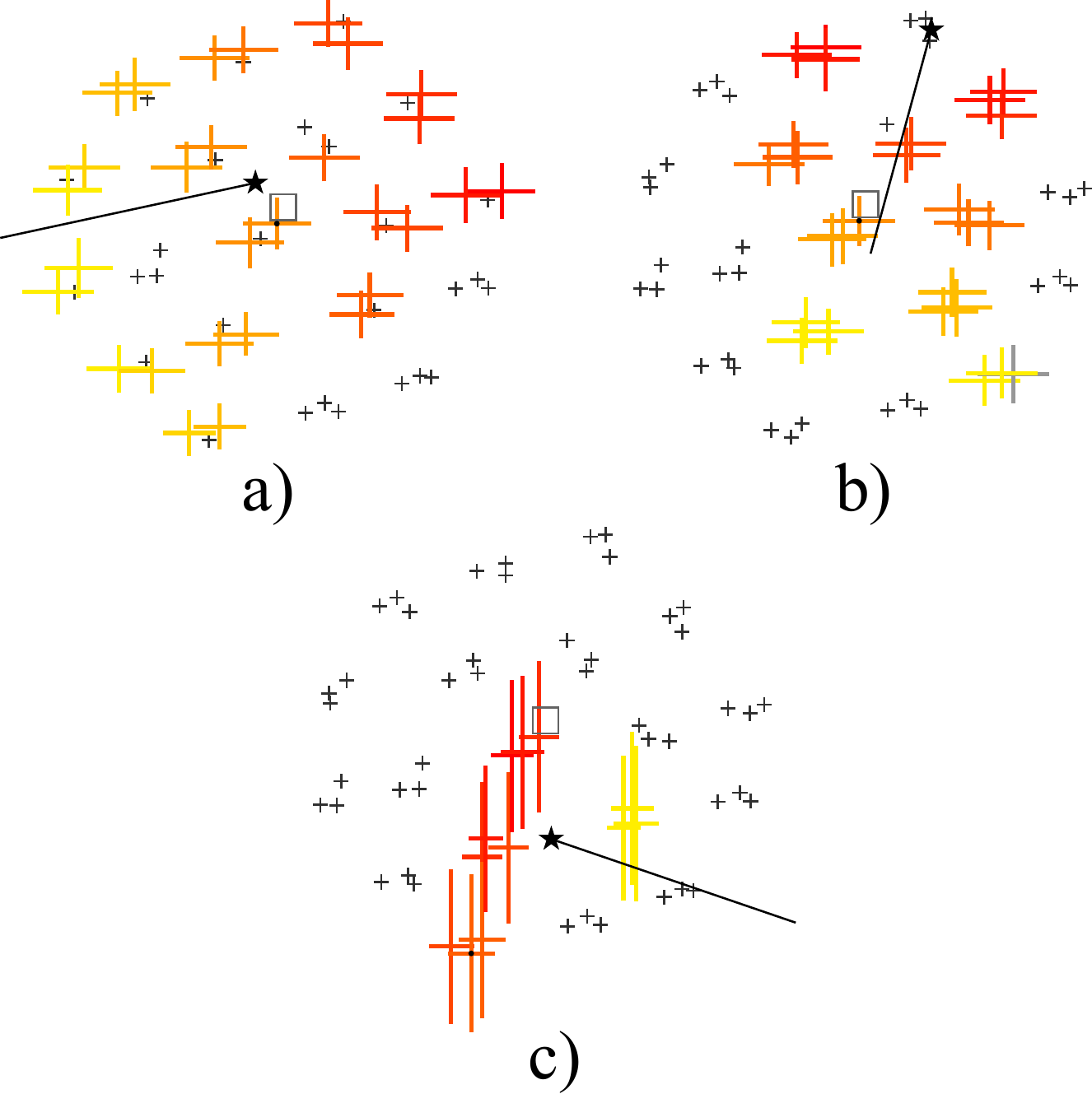}
	\caption{\textit{Left}: Rough energy estimation for inclined events.
	\textit{Right}: Examples of inclined events in different zenith angle ranges: a) 
	$60^\circ-70^\circ$, b) $70^\circ-80^\circ$, c) $80^\circ-90^\circ$. 
	Arrows denote the direction of the air-shower.
	Color code of crosses corresponds to the arrival time and their size indicates the signal strength (small ones are the antenna stations which were not read out).}
	\label{fig:inclined}       
\end{figure}

\section{Developing techniques for the next-generation detectors}

Tunka-Rex has a fruitful cooperation with the most of the modern digital radio arrays.
Besides cross-check of the energy scale performed with LOPES experiment, we have performed a cross-check of calibration using different methods applied to LOFAR telescope~\cite{Mulrey:2019vtz}.
Since many years, we are using and contributing to the radio extension of the Auger Offline analysis framework~\cite{Abreu:2011fb}.
The success of the SALLA instrument and corresponding electronics in the frame of Tunka-Rex experiment led to the choose of this instruments for the large radio array at the Pierre Auger Observatory~\cite{Hoerandel_ARENA2018}.
\label{sec:dev}

\textbf{Detection of inclined air-showers.}
The design of Tunka-Rex antennas features decreased sensitivity in lower hemisphere, which
prevents detection of signals reflected from the ground and reduces systematic uncertainties. However, the sensitivity to inclined events is reduced at the same time, which shifts the threshold by about one order of magnitude to the EeV range. 
Moreover, the sensitivity to inclined events is suppressed by the acceptance of the trigger. 
Nevertheless, we selected few high-energy events and performed a simple analysis of them.
One can see the results of the analysis in Fig.~\ref{fig:inclined}.
Despite of the small number of those events we confirmed that Tunka-Rex equipped with SALLA is sensitive for this type of events. 
Due to smeared LDFs, standard methods of reconstruction are not efficient and have low performance. 
Therefore, more sophisticated methods are required for the reconstruction of very inclined events.
The detailed description of this study can be found in ARENA2018 slides and proceedings~\cite{Marshalkina:2018sfx}, and we plan to continue our study due to high demand by future potential neutrino detection using this technique.

\textbf{Radio self-trigger} is the main challenge for the next-generation radio arrays.
On the one hand the radio background in the most locations is very challenging for the simple threshold triggers, on the other hand the modern electronics allows one to perform sophisticated computations and filtering on the fly.
We have started a research and development project which will result in the implementation of software for the modern FPGA and DAQ boards.
Our plan is to investigate possible trigger modes, implement and test the best one using a FPGA virtual machine, simulations, and TRVO data.

\begin{figure}[t!]
	\centering
		\includegraphics[width=0.67\linewidth]{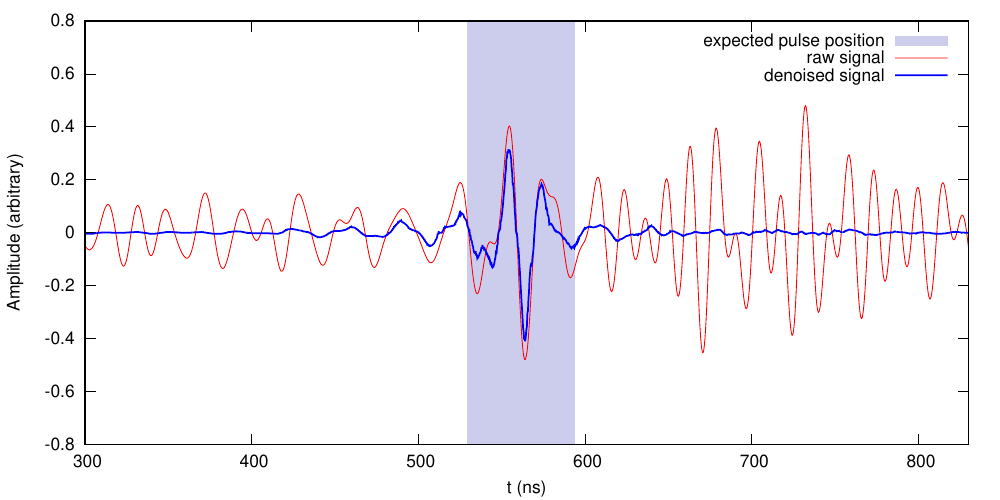}
		\caption{
			Example of the autoencoder performance on a measured Tunka-Rex trace showing successful denoising of the typical RFI after the signal.
		}
		\label{fig:real}
\end{figure}

\textbf{Deep learning for the signal denoising} is actively developed for the Tunka-Rex data.
We use autoencoder architecture with 1D convolutional layers, and have successfully proven that such a neural network can reconstruct simulated pulses even at low signal-to-noise ratios~\cite{Bezyazeekov:2019jbe}.
For the moment we switch to the application of this novel method for the reconstruction of real data, the example of which can be seen in Fig.~\ref{fig:real}.

\textbf{Tunka-Rex based pathfinder arrays.}
We have deployed few engineering arrays based on our antennas and electronics.
To test the air-shower emission in the high-mountain environment we have deployed the small air-shower array consisting of four SALLAs, which is the highest (3300~m a.s.l.) radio air-shower array~\cite{Beisenova:2017knp}\footnote{The website of the project \href{http://almarac.astroparticle.online}{http://almarac.astroparticle.online} is under development.}.
Another engineering array, namely Tunka-21cm~\cite{Tunka21cm_ICRC2019} is testing the possibility of application of air-shower hardware to the experiments aimed at the detection of weak signal from neutral hydrogen emitted and rescattered at redshifts of $Z>10$, since required frequency band and timing and spectral resolutions are close to the ones used in cosmic-ray science.

\section{Conclusion}
Tunka Radio Extension has grown from a small engineering array to the precise sparse radio detector with resolution compared to air-Cherenkov and fluorescence detectors during the seven years of its operation.
We have shown that the cost-effective radio detector maintained by the small team can be operated extremely efficiently.
Besides obvious contribution to the field, which is now succeeded by the next-generation large-scale detectors, the generation of radio experts from bachelors to professors were grown with help of our project.
The legacy of the detector will be held in the Baikal Open Laboratory for Astroparticle Physics~\cite{Bezyazeekov:2019pii}.

In 2019 we have performed the decommission of the Tunka-Rex detector, leaving only about 20\% antenna stations in operation for diagnostics and educational purposes.
These 20\% are antenna stations which have never been repaired and survived in the initial state during seven years.
We plan to prepare performance and aging report, which can be useful for the future experiments.

\section*{Acknowledgements}
\begin{small}
The construction of Tunka-Rex was funded by the German Helmholtz association and the
Russian Foundation for Basic Research (Grant No. HRJRG-303).
In preparation
of this work we used calculations performed on the HPC-cluster Academician V.M. Matrosov~\cite{HPC_Matrosov}
and on the computational resource ForHLR II funded by the Ministry of Science, Research
and the Arts Baden-W\"urttemberg and DFG (``Deutsche Forschungsgemeinschaft''). 
This work has been supported 
by the Russian Foundation for Basic Research (grants No. 18-32-00460 and No. 18-32-20220),
by the Russian Science Foundation Grant 19-72-00010 (section~\ref{sec:dev})
and by the Helmholtz grant HRSF-0027.
We also thank our former co-authors who made essential contributions to the Tunka-Rex development.
\end{small}
\bibliographystyle{ieeetr}
\bibliography{references}

\end{document}